\documentstyle [titlepage,12pt]{article}
\input{psfig}
\def\beq{\begin{equation}}
\def\eeq{\end{equation}}
\def\om{i\omega}
\def\R{{\bf R}}
\def\frac#1#2{{#1 \over #2}}
\def\lep{\left(}
\def\rip{\right)}
\def\leb{\left[}
\def\rib{\right]}

\def\etal{ et al.}

\def\t1{t_1}

\def\eq#1{(#1)}
\def\eqs#1#2{(#1)-(#2)}
\def\D{\displaystyle}

\title{\thispagestyle{myheadings}%
Magnetic and Dynamic Properties of the Hubbard Model in Infinite Dimensions}
\author{Mark Jarrell\\
        Department of Physics\\
        University of Cincinnati \\
        Cincinnati, Ohio 45221\\
	and \\
 	Thomas Pruschke\\
	Department of Physics\\
	The Ohio State University\\
	Columbus, Ohio  43210}

\date{\small May, 1992}

\begin{document}
\maketitle
\begin{abstract}
An essentially exact solution of the infinite dimensional Hubbard
model is made possible by using a self-consistent mapping of the Hubbard
model in this limit to an effective single impurity Anderson model. Solving
the latter with quantum  Monte Carlo procedures enables us to obtain
exact results for the one and two-particle properties of the
infinite dimensional Hubbard model. In particular we find antiferromagnetism
and a pseudogap in the single-particle density of states  for
sufficiently large values of the intrasite Coulomb interaction at half filling.
Both the antiferromagnetic phase and the insulating phase above the
N\'eel temperature are found to be quickly suppressed on doping. The
latter is replaced by a heavy electron metal with a quasiparticle mass strongly
dependent on doping as soon as $n<1$. At half filling the antiferromagnetic
phase boundary agrees surprisingly well in shape and order of magnitude with
results for the three dimensional Hubbard model.
\\

PACS numbers: 71.10+x, 75.10.Jm, 75.10.Lp, 75.30.Kz.

\end{abstract}

\section{Introduction}
The Hubbard model \cite{hubbard} of strongly correlated electron systems has
been an enduring problem in condensed matter physics.  It is
believed to at least qualitatively describe some of the properties of
transition
metal oxides, and possibly high temperature superconductors \cite{anderson-87}.
Using standard notation, the Hubbard Hamiltonian reads \cite{hubbard}
\beq
{\cal H} =\begin{array}[t]{l}
\displaystyle -t \sum_{<ij>,\sigma} \lep C^{\dagger}_{i,\sigma}
 C_{j,\sigma} + C^{\dagger}_{j,\sigma}
 C_{i,\sigma}\rip \\[5mm]
\displaystyle+
\sum_{i}\leb \epsilon \lep n_{i,\uparrow} + n_{i,\downarrow}\rip
+U \lep n_{i,\uparrow} -1/2 \rip \lep n_{i,\downarrow}- 1/2 \rip\rib
\,
\end{array}
\eeq
where $C_{i,\sigma}$ ($C^{\dagger}_{i,\sigma}$) destroys (creates) an
electron of spin $\sigma$ on site $i$ of a hypercubic lattice of dimension $d$,
and  $n_{i,\sigma}=C^{\dagger}_{i,\sigma}C_{i,\sigma}$.
Despite the simplicity of the model, no exact solutions exist
except in one dimension, where the knowledge is in fact
rather complete \cite{oned}. The unusual properties in one dimension
also gave rise
to discussions of whether the behaviour in 2D could be related to
the 1D case \cite{anderson-90}. The extent that this very special limit
can serve as a reference point for any finite dimensional model
\cite{schulz,anderson-92} remains controversial,
and an examination of the model from a different point of view is clearly
needed.

Recently, a new approach \cite{metzvoll,emh,dongen}
based on an expansion in $1/d$ about the point $d=\infty$ has been proposed to
study such strongly
correlated lattice models.  In this limit the requirement of a finite total
energy per site makes it necessary to rescale nonlocal interactions by an
appropriate power of $d^{-1}$ \cite{metzvoll,emh}. While e.g.\ spin-exchange
will essentially reduce to the corresponding mean field theory results
\cite{emh},
interactions like the screened Coulomb repulsion in the model (1) remain
nontrivial even in this limit.

In a previous short paper \cite{jarrell}, one of us presented the first
method of
exact solution of the Hubbard model in the infinite
dimensional limit.  In this paper, we provide a greatly expanded
discussion of the method and the physics.  The model in the infinite
dimensional limit retains the physics expected in the low dimensional model,
e.g.\ antiferromagnetism and the formation of a correlation induced
Mott-Hubbard pseudogap in the single-particle density of states. In addition,
we find some very interesting structures in the single-particle
density of states
(DOS), which we believe will persist in $d<\infty$ and lead to well defined
and observable effects, which may serve as a guide whether the model is
indeed capable of describing the essentials of transition metal physics.

\section{Theory and Numerical Considerations}
The limit as $d\to\infty$ is taken such that $4t^2d={t^*}^2={\rm{constant}}$
and it is convenient to choose $t^*=1$ as the energy scale for
the remainder of this paper.
This rescaling of the kinetic energy automatically leads to a finite
net effective magnetic exchange $d J\sim {t^*}^2/U$.
Generally this limit
provides two large simplifications.  First, the free DOS for near-neighbor
transfer
along the coordinate axis can be shown to acquire a Gaussian form, i.e.
\cite{metzvoll,emh}
\beq
\rho_0(t)=\exp(-t^2)/\sqrt{\pi}\;\;.
\eeq
Second, the problem reduces essentially to a local
one since nonlocal dynamical interactions become negligible in this limit
\cite{metzvoll,emh}. For example, the irreducible self energy and the
irreducible vertex functions purely are local.

This may be seen from a diagrammatic argument
\cite{schczy}. Consider the first few diagrams of the single-particle
self-energy  for this problem as shown in Fig.~1.
\begin{figure}[htb]
\centerline{\psfig{figure=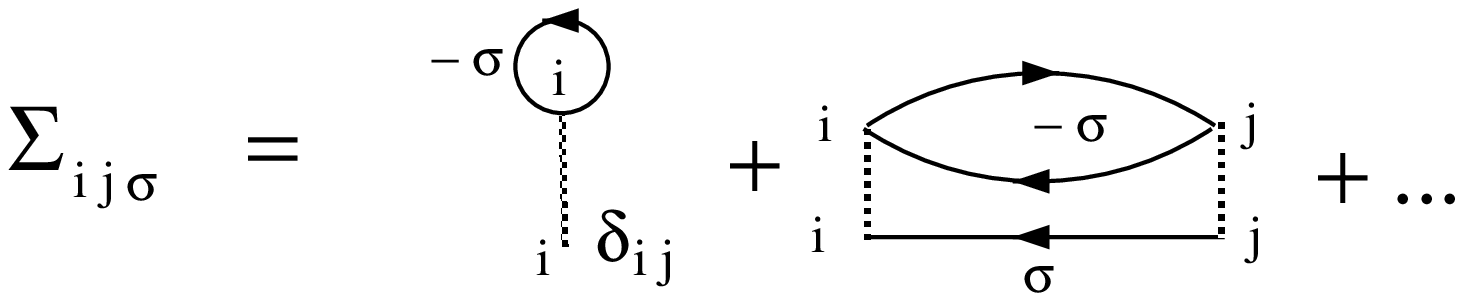,height=1.2in,width=7.5in}}

\caption[]{\em{First few diagrams for the lattice self energy.
Here, the solid lines represent the undressed ($U=0$) electron
propagators $G_{ij}^0(\om_n)$, and the dashed lines represent the local
Coulomb interaction $U$.}}
\end{figure}
This is a real-space representation, so each electron propagator
$G_{ij}$ scales as $\sim t^{-|\R_i-\R_j|}$.
Thus the second order term in Fig.~1 scales as $ t^{-3|\R_i-\R_j|}$ and,
after summing over the contribution of the nearest neighbor shell, gives a
contribution of the order $d\cdot d^{-3/2}$ (since $4t^2d=1$).
This contribution vanishes in the limit as $d\to \infty$. A similar
argument may be applied to all terms, and only the
site diagonal self-energy survives the limit $d\to\infty$.
Furthermore, since the lattice is translationally invariant
$\Sigma_{ij}(\om_n)=\Sigma(\om_n)\delta_{ij}$ independent of $i$.
Thus, Dyson's equation reduces to
\beq
G_{ij}(\om_n)=G_{ij}^0(\om_n) +\sum_k G_{ik}^0(\om_n) \Sigma(\om_n)
G_{kj}(\om_n)
\eeq
with the diagrammatic equation for $\Sigma(z)$ in Fig.~1. A completely
equivalent
argument may be used for the magnetic susceptibility \cite{hz}, which reads
\begin{equation}
\begin{array}{l@{\;=\;}l}
\displaystyle
\chi_{ij}(i\nu_n)&\displaystyle\frac{1}{\beta}
\sum_{\omega_n,\omega_m}\tilde\chi_{ij}(i\omega_n,i\omega_m;i\nu_n)\\[5mm]
\displaystyle\tilde\chi_{ij}(\om_n,\om_m;i\nu_n)&
\begin{array}[t]{l}
\displaystyle\tilde\chi^0_{ij}(i\omega_n;i\nu_n)\delta_{nm}\\[3.5mm]
\D+\frac{1}{\beta}
\sum_{\omega_p,l}\tilde\chi^0_{il}(i\omega_n;i\nu_n)\Gamma(i\omega_n,i\omega_p;
i\nu_n)\tilde\chi_{lj}(i\omega_p,\om_m;i\nu_n)\end{array}\\[20mm]
\displaystyle\tilde\chi^0_{ij}(i\omega_n;i\nu_n) & \D\frac{1}{N^2}
\sum_{{\bf k},{\bf q}}e^{i{\bf q}\cdot({\bf R}_i-{\bf R}_j)}
G_{\bf k}(i\omega_n)G_{{\bf k}+{\bf q}}(i\omega_n+i\nu_n)
\end{array}
\eeq
where $\beta=(k_BT)^{-1}$, $\omega_n$ ($\nu_n$) are the Fermi- (Bose-)
Matsubara
frequencies, $G_{\bf k}(z)$ is the one particle Green's function obtained from
(3) and $\Gamma(z,z')$ the irreducible vertex.

Usually, quantities like the self-energy $\Sigma(z)$ and $\Gamma(z,z')$
have to be evaluated using a direct perturbation expansion as outlined in
Fig.~1. With respect to $U$ this has been done for the present model quite
extensively by means
of conserving approximations including particle-hole and particle-particle
ladder summations \cite{menemh}. Alternatively, one of us recently proposed a
perturbation theory using the transfer term $t$ as expansion
parameter \cite{pr90}.
However, as one knows from similar studies for
heavy fermion systems, the validity and radius of convergence of these
expansions
may depend critically on the particular choice of parameters \cite{zlahor}.
In the following we therefore want to present an approach which maps the
whole problem to the solution a single impurity Anderson model, where
several reliable ways to obtain the solution are known.
That this connection exists was realized quite early \cite{bramil}, but
until now it only served to set up equations for the thermodynamic potential
\cite{bramil,janis}.

To do this let us return to equation (3). For $i=j$, the diagrams contributing
to the
local Green's function ${\cal G}\equiv G_{ii}$ may be rearranged to contain
explicitly only local processes as illustrated in Fig.~2.
\begin{figure}[htb]
\centerline{\psfig{figure=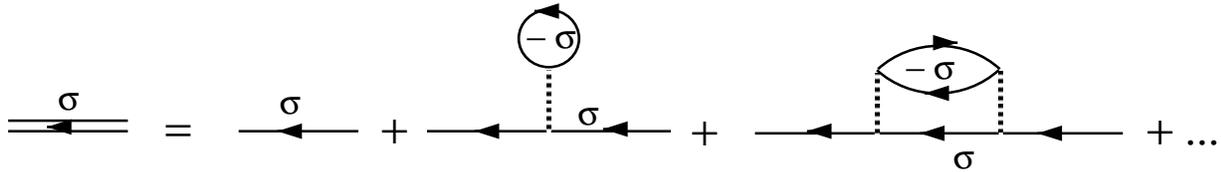,height=1.25in,width=7.5in}}

\caption[]{\em{First few diagrams for the local Green's function
$\cal G$ (double solid line). The diagrams have been rearranged such that all
processes occurring on other sites than the one considered were included into
an
effective host Green's function $\tilde{\cal G}$ represented here as a
solid line. The Coulomb interacting is again visualized as a dashed line.}}
\end{figure}
The undressed Green's function represented by the solid line in
Fig.~2 must incorporate the missing nonlocal
processes and thus is the solution to a modified lattice problem
\beq
\tilde{\cal G}(\om_n)=\tilde{G}_{ii}(\om_n)= {G}_{ii}^0(\om_n) +
{\sum_k}' {G}_{ik}^0(\om_n)\Sigma(\om_n) \tilde{G}_{ki}(\om_n)
\eeq
where the prime at the sum indicates that $k\ne i$ and $G_{ij}^0(z)$ is the
unperturbed propagator. This summation restriction
is necessary to avoid  over-counting of local diagrams. In reciprocal space,
equation (5) reads
\beq
\tilde{G}_{\bf k}(\om_n)= {G}_{\bf k}^0(\om_n) +
{G}_{\bf k}^0(\om_n)\Sigma(\om_n)\left(\tilde{G}_{\bf k}(\om_n)-\tilde{\cal G}(
\om_n)\right)\;\;.
\eeq
This may be solved for $\tilde{G}_{\bf k}(z)$ and summed on $\bf k$ to yield
\beq
\frac{1}{\tilde{\cal G}(z)}=\frac{1}{{\cal G}(z)}+\Sigma(z)\;\;.
\eeq

The problem closes by noting two things. First, Fig.~2 is nothing but the
perturbation expansion of an Anderson impurity model with a host Green's
function specified by $\tilde{\cal G}$.
In the spirit of this analogy, one may then define an effective hybridization
strength $\Delta(z)$ via \cite{c1}
\beq
\Delta(z)=\frac{1}{\tilde{\cal G}(z)}-z+\epsilon\;\;.
\eeq
Secondly, $\Sigma(z)$ and $G_{ii}(z)$ are as usual related by
\beq
G_{ii}(z)=\int d\omega\rho_0(\omega)\frac{1}{z-\omega-\epsilon-\Sigma(z)}
\;\;.
\eeq
However, $G_{ii}(z)$ is by definition also the propagator of the effective
Anderson
impurity problem {\em defined}\/ by equation \eq{8}.
Let us emphasize, that, although we used the formal perturbation
expansion of the model (1) in terms of $U$ in our discussion, it does not
enter the set of equations \eqs{5}{9} at any stage. This means
that we did not refer to any special method to solve the impurity Anderson
model and that equations \eqs{5}{9} are exact.

In the present paper we should like to concentrate on the
quantum Monte Carlo (QMC) algorithm of
Hirsch and Fye \cite{hirsch,mj1} as a method to solve the impurity Anderson
model, although equivalent calculations were
done using a self-consistent perturbation scheme (NCA) for this problem
\cite{prgr89}. These calculations generally use considerably less computational
time and, away from some critical region at half filling, the
results found were in good agreement with the QMC \cite{mpp}.
However, with respect to  two-particle properties the QMC is simply the
more adequate approach, since the NCA requires the solution of difficult
Bethe-Salpeter type equations as soon as finite Coulomb repulsion is
considered.

In the QMC the problem is cast into a discrete path formalism in imaginary
time, $\tau_l$, where $\tau_l=l\Delta\tau$, $\Delta\tau=\beta/L$,
and $L$ is the number of times slices.  The values of $L$ used
ranged from $40$ to $160$, with the largest values of $L$ reserved for
the largest values of $\beta$  since the time required by the algorithm
increases like $L^3$.  No sign problem was encountered in the QMC process,
except for extremely large values of $U$ away from half filling.
Even here, the sign problem was mild, and easily handled with the
standard methods.
The self-consistency process described above is easily employed
in the QMC process. Choosing an intial $\Sigma$, the output of the process
is ${\cal G}$, which may then be inverted to yield a new estimate for
$\Sigma(\om_n)$ and so on until ${\cal G}=G_{ii}$ within the numerical
precision.  Usually 4--8 iterations are required for convergence.  Other
quantities such as the unscreened local moment
$\mu^2=<(n_{\uparrow}-n_{\downarrow})^2>$, the two-particle Green's functions
equation (4) etc. are calculated on the last iteration, once convergence is
reached.  Typical systematic (due to finite $\Delta\tau$) and statistical
errors were quite small, usually less than 1\%.  Thus, unless explicitely
displayed, the error bars on the static quantities may be assumed
to be smaller than the plotting symbols.

\section{Two-Particle Properties}

As already mentioned, a variety of two-particle properties may be
calculated with this procedure. Here we want to concentrate on the
antiferromagnetic static susceptibility, given by (4) for
$i\nu\to0$ at ${\bf q}={\bf Q}=(\pi,\pi,\cdots)$. Except for
the one particle propagators discussed in the next section, the only
unknown quantity there is the local vertex $\Gamma$. However, because it
involves only local processes, it can be determined from the susceptibility
of the impurity Anderson model, which may be obtained from
\beq
\tilde{\chi}_{ii}(\om_n,\om_m;i\nu_n)=\begin{array}[t]{l}
\D  \tilde{\chi}_{ii}^0(\om_n;i\nu_n)
\delta_{nm} \\[5mm]
\D+
\frac{1}{\beta} \sum_p  \tilde{\chi}_{ii}^0(\om_n;i\nu_n)
\Gamma(\om_n,\om_p;i\nu_n)
\tilde{\chi}_{ii}(\om_p,\om_m;i\nu_n)
\end{array}
\eeq
Here $\tilde{\chi}_{ii}$ is the opposite-spin two-particle Green's function
matrix,
\beq
\tilde{\chi}_{ii}(\om_n,\om_m;i\nu_n)= -\frac{1}{\beta^2} \int_0^\beta
\begin{array}[t]{l}
\D d\tau_1 \int_0^\beta d\tau_2
\int_0^\beta d\tau_3\int_0^\beta d\tau_4
\displaystyle e^{-\om_m(\tau_1-\tau_2)}
e^{-\om_n(\tau_3-\tau_4)}\\[5mm]
\displaystyle e^{i\nu_n(\tau_2-\tau_4)}\left< T_\tau C_{i,\uparrow}(\tau_4)
C^{\dagger}_{i,\downarrow}(\tau_3)C_{i,\downarrow}(\tau_2)
C^{\dagger}_{i,\uparrow}(\tau_1) \right>\,.
\end{array}
\eeq
If we substitute \eq{10} into \eq{4}, we obtain for $i\nu\to0$ the matrix
relation
\beq
\tilde{\chi}_{\bf q}^{-1}=\tilde{\chi}_{\bf q}^{0\,-1}+
\tilde{\chi}_{ii}^{-1}-\tilde{\chi}_{ii}^{0\,-1}
\eeq
where the static susceptibility is obtained by summing
\beq
\chi_{\bf q}(T)=\frac{1}{\beta}\sum_{mn}\tilde{\chi}_{\bf q}(\om_n,\om_m)
\eeq

It is expected that the Hubbard model on a cubic lattice with nearest-neighbour
transfer will exhibit antiferromagnetism at half filling for arbitrary values
of $U$ due to perfect nesting \cite{lpm}.  This transition is signaled by the
divergence of the antiferromagnetic susceptibility $\chi_{AF}$
calculated using the methods described above.  Results from this approach
are shown in Fig.~3 for $U=1.5$ and $\epsilon=0$.
\begin{figure}[htb]
{\centerline{\psfig{figure=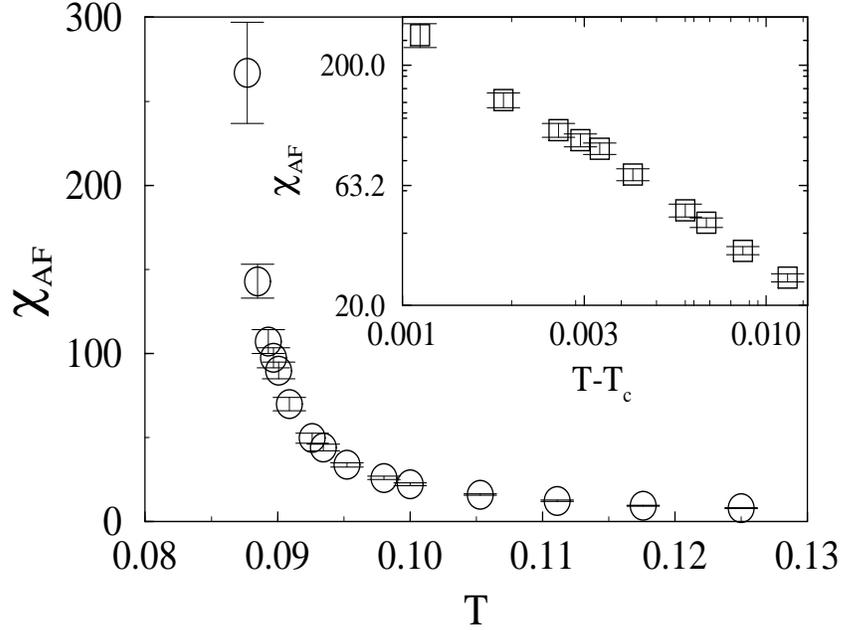,height=3.5in,width=4.5in}}}

\caption[]{\em{Antiferromagnetic susceptibility $\chi_{AF}(T)$,
versus temperature $T$ when $U=1.5$ and $\epsilon=0.0$.  The logarithmic
scaling behaviour is shown in the inset.  The data close to the
transition fit the form $\chi_{AF}\propto \left| T-T_N\right|^\nu$
with $T_N=0.0866\pm 0.0003$ and $\nu=-0.99 \pm 0.05$ consistent with the
mean-field behaviour expected for $d=\infty$.}}
\end{figure}
The logarithmic scaling behaviour is shown in the inset.
Near the Ne\'{e}l temperature
$T_N$ the data fit a form $\chi_{AF} \propto |T-T_N|^{\nu}$ with
$T_N=0.0866 \pm 0.0003$ and $\nu=-0.99 \pm 0.05$.  This scaling behaviour
is consistent
with that of a Heisenberg model on a lattice with an infinite
number of nearest neighbors, for which one expects the Curie-Weiss
mean-field form for $\chi_{AF}$.

The antiferromagnetic transition temperature $T_N$ for the half-filled model
\begin{figure}[htb]
\centerline{\psfig{figure=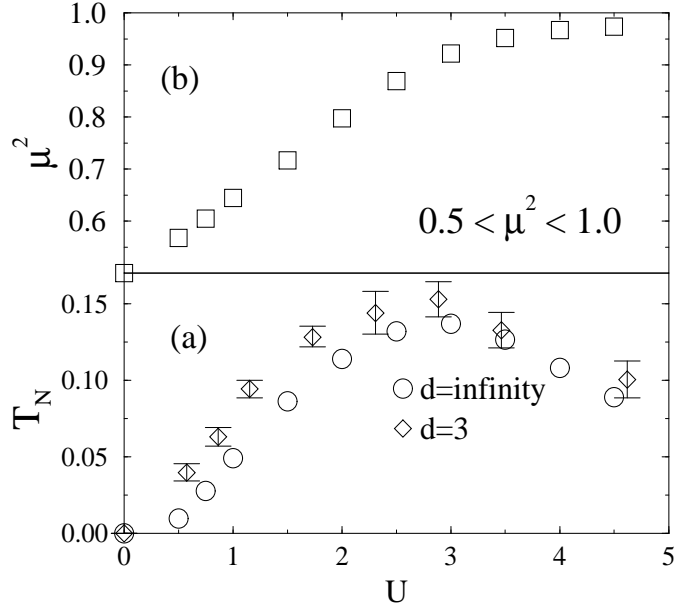,height=3.25in,width=4.25in}}

\caption[]{\em{(a) Antiferromagnetic $T_N$  and (b)
$\mu^2=<(n_{\uparrow}-n_{\downarrow})^2>$ (calculated at $T=T_N$)
as functions of $U$ for the half-filled model ($\epsilon=0$). In (a) the
circles
represent the values obtained from the present approach while the diamonds
are data for a three dimensional Hubbard model extracted from ref.~24.}}
\end{figure}
obtained within the current approach is plotted as a function of $U$ in
Fig.~4a.  For small values of
$U$, where the local spin moment is also small, we find that $T_N$
is exponentially small, consistent with
perturbation theory \cite{pgjvd}.  For very large values of $U$,
where the spin moment has saturated to its maximum value, one expects
that the transition temperature will fall monotonically with increasing $U$,
$T_N\sim1/U$ \cite{pgjvd}.  This is because the antiferromagnetic exchange
$d J\sim {t^*}^2/U$ also decreases with increasing $U$.
Thus, one expects a peak in $T_N(U)$ for some intermediate value of
$U$ as seen in Fig.~4a.  For comparison we included in Fig.~4a also $T_N(U)$
for the $d=3$ Hubbard model as calculated by Scalettar et al.\ \cite{scal}.
The shape and order of magnitude compare very well, although the $d=3$ data
always have slightly larger values, which we think is partially a
consequence of the
different analytic structures of the free DOS in $d=3$ and $d=\infty$
\cite{comment3}.
In Fig.~4b the unscreened
squared magnetic moment $\mu^2=<(n_{\uparrow}-n_{\downarrow})^2>$, calculated
at
the transition $T=T_N$, is plotted versus $U$ when $\epsilon=0$.  For the
half-filled model $\mu^2$ ranges from $\mu^2=0.5$ in the uncorrelated
\begin{figure}[htb]
\centerline{\psfig{figure=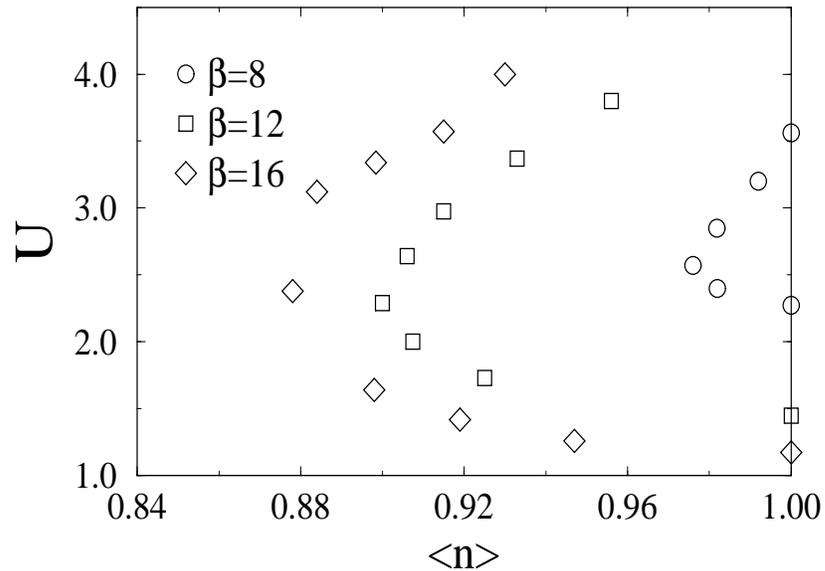,height=3.25in,width=4.25in}}

\caption[]{\em{Critical values of $U$, where $\chi_{AF}$ diverges,
versus filling for several temperatures.  The result is symmetric around
$<n>=1$.}}
\end{figure}
limit ($U=0$), to $\mu^2=1$ in the strongly correlated limit ($U\to\infty$).
Note that the peak in $T_N(U)$ occurs near the point where $\mu^2$ begins
to saturate to one.   Away from half filling, the divergence of the
antiferromagnetic susceptibility is quickly suppressed.  This behaviour
is shown in Fig.~5 where the critical value of $U$ is
plotted versus filling for several values of $\beta$.

In contrast to the antiferromagnetic susceptibility, the ferromagnetic
susceptibility (${\bf q}=(0,0,\cdots)$) is rather featureless
both as a function of temperature
\begin{figure}[htb]
\centerline{\psfig{figure=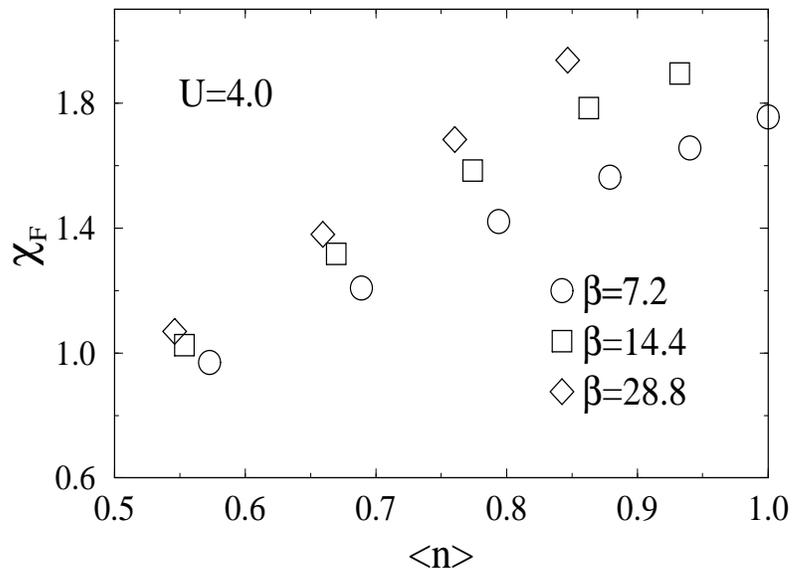,height=3.25in,width=4.25in}}
\caption[]{\em{The $\chi_F$ versus filling, when $U=4$. $\chi_F$ was not
calculated at half filling when $\beta=14.4$ and $\beta=28.8$ since
$\chi_{AF}$ has already diverged here. }}
\end{figure}
and filling.  This behaviour is shown in Fig.~6 where $\chi_{F}$ is
plotted versus filling for several temperatures when $U=4$.  For values of
$U$ as large as eight, the ferromagnetic susceptibility never diverged.
Special attention was placed near the region just off half filling for
large $U$ where one might expect Nagaoka behaviour.  In this region,
for $U=8$, $\chi_{F}$ did show a mild peak as a function
of filling.  However, this data was prone to excessive error bars, and
in all such cases $\chi_{AF}\gg\chi_{F}$.

In addition to ferromagnetic and antiferromagnetic behaviour, it is
possible that the infinite dimensional Hubbard model may exhibit incommensurate
order away from half filling.  Such calculations were not included here
but are left for future study.

	By a method very similar to that used to calculate the magnetic
susceptibilities, one may also calculate the superconducting pair-field
susceptibilities.  Since singlet superconducting order paramters with the
same symmetry (eg. s-wave and extended s-wave) mix, the possible
divergencies of all pair field susceptibilites with the same symmetry must
coincide.  Thus, it is only necessary to calculate one susceptibility of
each symmetry.  Furthermore, in the infinite dimensional limit, a pair-field
susceptibility which corresponds to a symmetry orthogonal to the lattice
symmetry (extended s-wave symmetry) has no vertex corrections.  Consider a
superconducting channel which has a form factor $t_{\bf k}$ such that
\beq
\sum_{\bf k} t_{\bf k} |G(\bf k,i\omega_n)|^2 =0\;\;.
\eeq
The first two diagrams in the corresponding pair-field susceptibility
are shown in figure~7.
\begin{figure}[htb]
\centerline{\psfig{figure=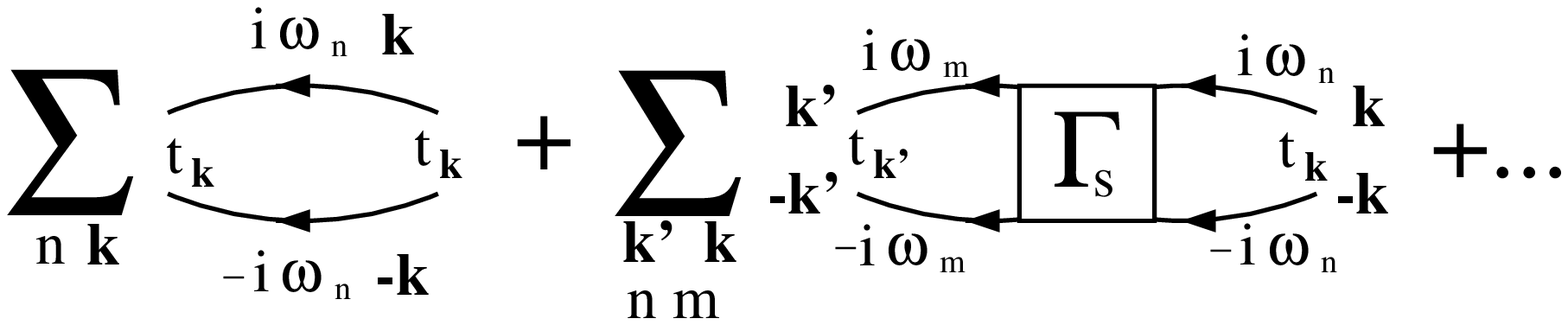,height=1.5in,width=6.375in}}
\caption[]{\em{First two diagrams for the superconducting pair-field
susceptibility.  The form factor $t_{\bf k}$ determines the symmetry of
the pairing.  If the symmetry of $t_{\bf k}$ is orthogonal to s-wave,
then the vertex corrections vanish.}}
\end{figure}
When $d=\infty$ the irreducible vertex has no k-dependence so the sum
over ${\bf k}$ may be performed independently
on the right and left-hand sides of the vertex in the second term.
Thus, when $t_{\bf k}$ is consistent with Eq.~14, this term vanishes.
The only diagram which remains is the simple bubble which can only
diverge in the zero-temperature limit.  This argument may be extended
to any superconducting order parameter which is orthogonal to s-wave, leaving
only the possiblity of superconductivity with s-wave symmetry.  However,
for all parameters dicussed here we found that the s-wave pair-susceptibility
remained finite at a value below the non-interacting limit.  Thus, we
conclude that there is no singlet superconductivity in the infinite
dimensional Hubbard model.

\section{Single-Particle Properties}

In this section we will discuss the single-particle properties of the
Hubbard model (1) as $d=\infty$. The most important quantity in this
connection is the single particle density of states defined by
$A(\omega)=-1/\pi {\rm{Im}} G_{ii}(\omega +i0^+)$.
Since the QMC method supplies us
only with the imaginary time Green's function $G_{ii}(\tau)$, we used the
maximum entropy procedure \cite{method,bryan,rns} for providing its analytic
continuation. In most cases results from a self-consistent
perturbation theory for the Anderson impurity (NCA) \cite{prgr89,bcw} were used
as
a default model for this procedure.  This is a nearly ideal combination
of methods, since the NCA becomes exact at high frequencies where the
QMC data contains little information, and the analytically
continued QMC data is essentially exact at low frequencies. A detailed
comparison of NCA and QMC results will be presented elsewhere \cite{mpp}.
When a pseudogap was present in the one-particle excitation
spectrum, we found that the NCA equations failed to converge due
to numerical instabilities.
In this limit second-order perturbation theory in $U$ was used to provide
a default model for the maximum entropy process.

With either default model,
it was extremely important to carefully control both the systematic and
especially the statistical errors of the Monte Carlo process.  Thus,
before analytic continuation, the self consistent process discussed
above was allowed to fully converge.
Then several runs were performed to obtain a highly accurate average
value of $\Sigma$.  The self-consistency was then turned off, and
several runs using this value of $\Sigma$ were performed to calculate a
highly accurate value of ${\cal G}(\tau)$ which was then analytically
continued.

Results for the half-filled model are shown in Fig.~8.  Here $A(\omega)$ is
plotted for several values of $U$ when $\beta=7.2$, $\epsilon=0$,
and $<n>=1$.
\begin{figure}[htb]
\centerline{\psfig{figure=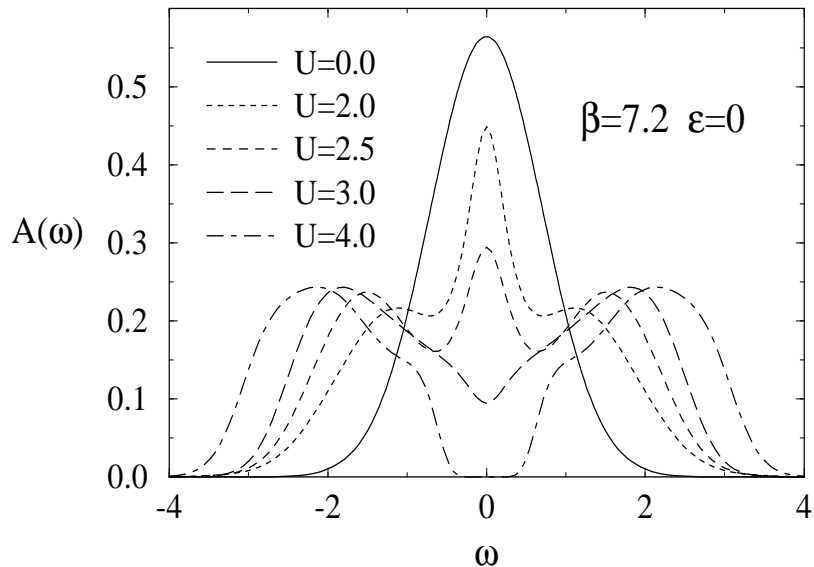,height=3.25in,width=4.25in}}

\caption[]{\em{Evolution of $A(\omega)$ in the paramagnetic
phase of the half-filled model.}}
\end{figure}
As the Coulomb repulsion $U$ is increased from zero the broad central peak
in the spectrum becomes narrower and is gradually suppressed while
at the same time two side bands build up which are carrying the majority of
the spectral weight. This kind of behaviour has also been found
in different kinds of perturbation theory \cite{pr90,georges}.
As discussed below, the central peak can
be identified with a Kondo-like resonance, and the side peaks as usual with
charge transfer on and
off the site.  As $U$ continues to rise, the spectrum begins to
develop a pseudogap at zero frequency when $U>3.4\approx U_C$.
The ``critical'' value $U_C$ increases with increasing temperature,
and for $U>U_C$ the pseudogap grows linearly in
$U$.  This structure is identified as a pseudogap, rather than a gap,
since the zero frequency
density of states was never identically zero.   Rather at $\omega=0$ it was
about 4 orders of magnitude smaller than that of the surrounding features.
Indeed, this is consistent with an argument of Krauth
and George \cite{krauth} who argue that in this model the zero frequency
density
of states is finite for all $T$, even at $T=0$.

The evolution of density of states and the filling as a function of $\epsilon$
is shown in figure 9.  As $\epsilon$ is increased from zero, the
filling decreases very slowly at first, consistent with the presence of a
pseudogap.  The filling begins to change rapidly once $\epsilon$ is
increased to roughly half the gapwidth.  The change in filling is
enhanced by the presence of a narrow resonance near zero frequency
which replaces the pseudogap as soon as the model is doped away from
half filling.
\begin{figure}[htb]
\centerline{\psfig{figure=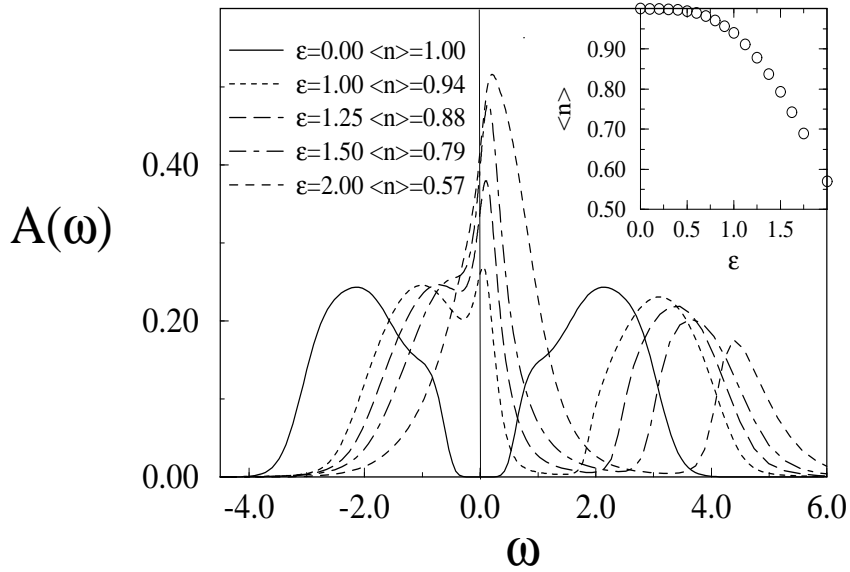,height=3.25in,width=4.25in}}

\caption[]{\em{Evolution of $A(\omega)$ as a function of filling
when $\beta=7.2$ and $U=4$.  The filling versus $\epsilon$ is shown
in the inset.}}
\end{figure}
However, there is a vestige of the Mott pseudogap at higher frequencies.

Perhaps even more interesting is the temperature dependence of
the DOS. In Fig.~10 the DOS
\begin{figure}[htb]
\centerline{\psfig{figure=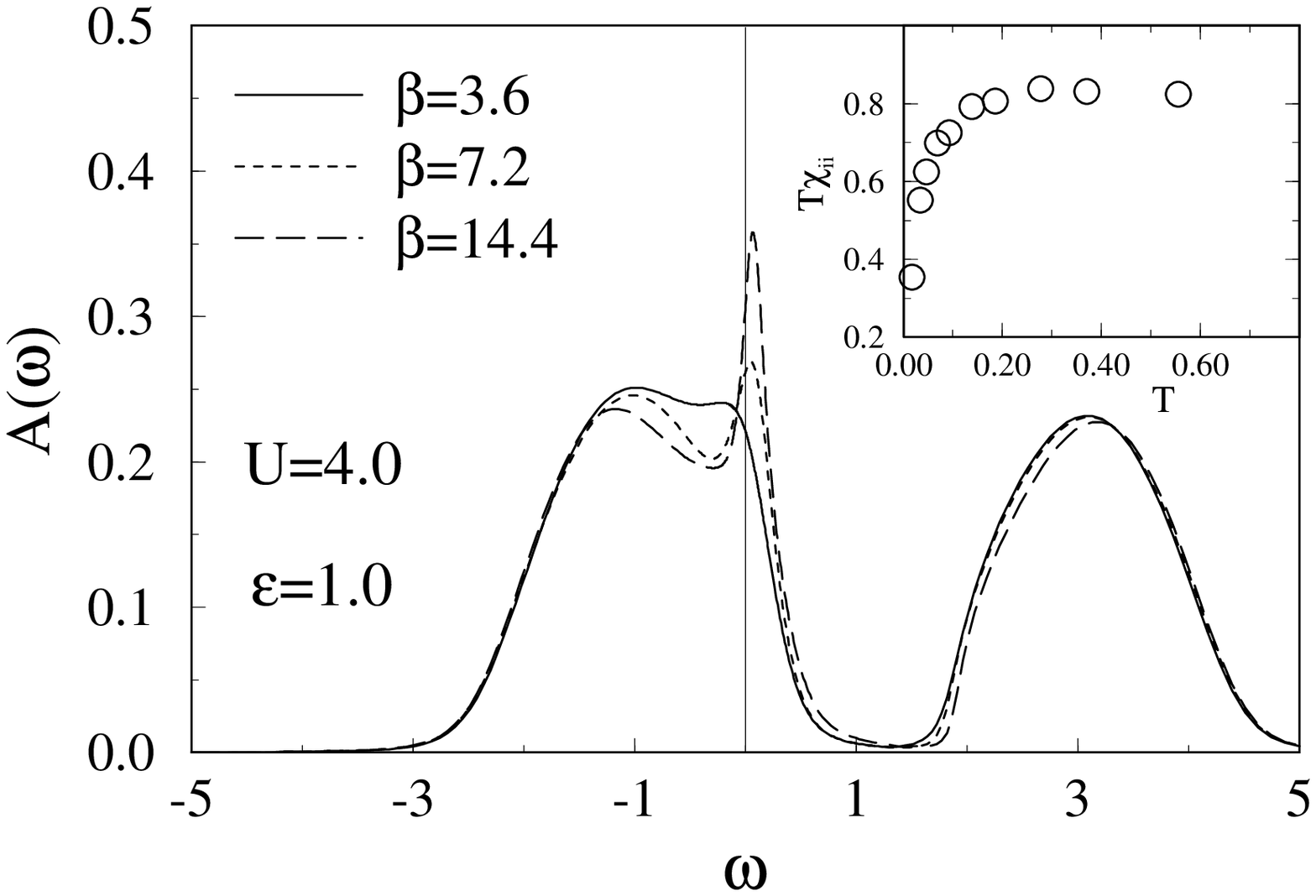,height=3.0in,width=4.0in}}

\caption[]{\em{Evolution of the Kondo-like feature with $T$.
As shown in the inset,the growth of the resonance is correlated with the
reduction of the screened local moment $T\chi_{ii}$.}}
\end{figure}
for $U=4$ close to half filling ($n\approx0.94$) is displayed for some typical
values
of $T$. While the peaks related to charge excitations are nearly temperature
independent, the resonance at the chemical potential
is very sensitive to the temperature and evolves roughly as $\ln(T)$ when going
from $\beta=7.2$ to $\beta=14.4$.
It becomes more pronounced at lower temperatures,
and apparently  disappears completely once the temperature is raised
to about twice the width of this feature.
As shown in the inset in Fig.~10, the growth of this resonance
is correlated with the reduction of the screened local
moment $T\chi_{ii}(T)$. In connection with the fact that eqs. \eqs{5}{9}
represent an effective Anderson impurity problem, it is evident that this
behaviour must be interpreted as a Kondo-like screening of the local moment
and consequently the resonance at the Fermi surface
can be identified with the Abrikosov-Suhl
resonance known to accompany the former.
With the reasonable assumption $\Sigma(i0^+)
\to0$ as $T\to0$, we expect $A(0)\to1/\sqrt{\pi}$ \cite{emuha2}. Using as
the ``Kondo''-scale the temperature, where $A(0;T_K)=\frac{1}{2}A(0,T\to0)$, we
find $T_K\approx t^*/8$ in this case.

Since this Kondo-like feature is pronounced, and occurs right
near the Fermi surface, one would expect it to have a strong effect
upon a number of properties of the system like resistivity and specific
heat \cite{mpp}. One simple way to obtain an idea how large this influence
may be is to inspect the effective mass of the quasiparticles in the system.
In Fig.~11, the mass enhancement factor
\beq
a^{-1}(T)=1-\lep {\rm{Im}}\Sigma(\om_0)\rip/\omega_0
\eeq
is plotted versus temperature when $U=4$ for some typical values of $\epsilon$.
\begin{figure}[htb]
\centerline{\psfig{figure=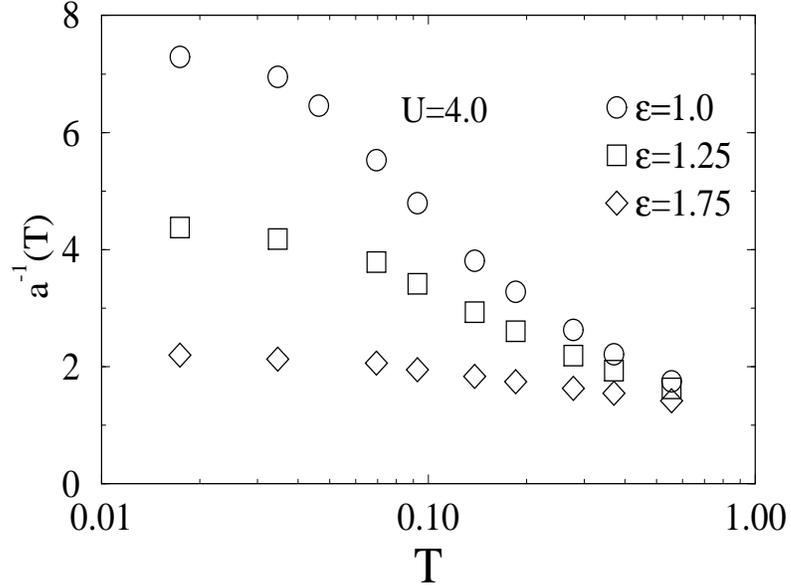,height=3.25in,width=4.25in}}

\caption[]{\em{Mass enhancement
$a^{-1}(T)=1-\lep{\rm{Im}}\Sigma(\om_0)\rip/\omega_0$ versus temperature.}}
\end{figure}
Here, $\om_0$ is the lowest Matsubara frequency, and the wave function
renormalization factor is given by the zero temperature limit
of $a(T)$.  As shown in the figure, the mass enhancement factor increases
like $\ln(T)$ until the temperature falls well below the
Kondo-scale, in which case it saturates.
The behaviour $a^{-1}(T)\sim\ln(T)$ has been
identified \cite{serene} as one of
the signatures of marginal Fermi liquid behaviour which has
been suggested as a possible phenomenological picture for the two-dimensional
limit of the
model (1) \cite{ffnj}. However, it is clear that the infinite-dimensional
model close to half filling is not marginal, since $a^{-1}$ saturates to a
finite
value at low temperatures, corresponding to a usual Fermi liquid with
an enhanced effective
mass, which is strongly dependent upon filling.
In the case when $\epsilon=1.0$, the filling is roughly $<n>\approx 0.94$,
and $a^{-1}(0)\approx 7.5$, corresponding to a mass enhancement of
$m^*/m\approx 7.5\approx T_K^{-1}$.
Far away from half-filling, it is about one, and as half filling
is approached it increases dramatically.

\section{Conclusion and Interpretation}

The method described in this paper has reduced the infinite dimensional
Hubbard model to a self-consistently embedded Anderson impurity
problem.  Thus for large $U$ the qualitative features of the density
of states have an obvious interpretation in terms of the Anderson model
spectrum.  The upper and lower peaks found correspond to charge transfer on
and off the local site. In addition a small resonance at or near the chemical
potential occurs
which to our opinion must be interpreted in terms of an
Abrikosov-Suhl resonance accompanying the screening of the local moment as
observed in $T\chi_{ii}(T)$. As well known from the physics of the Kondo
effect,
this formation of a bound state causes a significant mass enhancement
for the quasiparticles in the system as $T\to 0$.

	From the data presented here, it is possible to obtain
a very good representation of the phase diagram of the infinite dimensional
Hubbard model.  For example, the phase diagram for half-filling is
illustrated in Fig.~12.
\begin{figure}[htb]
\centerline{\psfig{figure=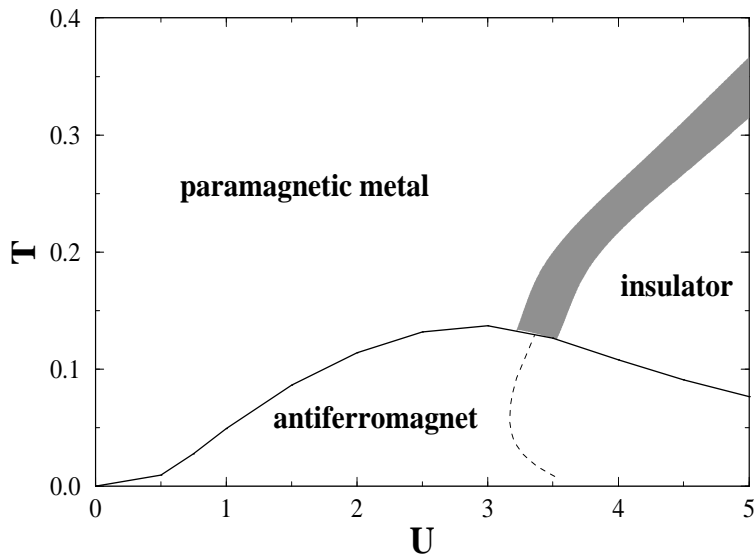,height=3.25in,width=4.25in}}

\caption[]{\em{Phase diagram for the Hubbard model in $d=\infty$
at half filling. The shaded region is centered upon the location
of the continuous metal-insulator transition.  The extension of this
boundary into the antiferromagnetic region is represented by the
dashed line.}}
\end{figure}
\noindent At temperatures above $T_N$ we find a paramagnetic Fermi liquid state
with more or less enhanced quasiparticle masses. This evolves gradually into
a Mott-Hubbard phase which is stable beyond some critical value of $U$.
This ``insulator'' phase is identified with
the occurance of an exponentially small density of states at zero
frequency and is quickly suppressed upon increase of $T$, because it can only
be stabilised when the charge excitation peaks are well separated on the scale
of variation of Fermi's function.  Since the zero frequency density of states
is always finite,
no real phase transition (i.e. one with an identifiable order parameter)
from the metal to the insulator has occurred.  However, when the
zero-frequency density of states becomes exponentially small, the dynamic
and and transport
properties \cite{mpp} of the system change radically in a manner
consistent with the change from a metal to an insulator. At low temperatures
the system is always antiferromagnetically ordered with a symmetry-induced
gap in the one-particle DOS.
Upon doping (not shown),
the Mott insulator phase vanishes, and the AF region shrinks.

If we artificially
extend the metal-insulator boundary into the antiferromagnetic
region (the dashed line in Fig.~12), then we find an interesting
reentrance: the critical $U_c$ increases
with decreasing temperature.
This reentrance is not relevant to the present unfrustrated model
due to the antiferromagnetism;
however, it is relevant to models with sufficient lattice frustration so
that the antiferromagnetic transition is suppressed.  Such is the case
if we include a hopping term to the second next neighbor along each of the
coordinate axes equal to the nearest neighbor hopping $t$.
In fact, with this modification, the unperturbed density
of states in the $d=\infty$ limit remains Gaussian, with a renormalized
energy scale \cite{emuha2}.  Our present calculation is relevant
to such a model. In our opinion this reentrance
must be attributed to a
competition between the Kondo-effect, which never completely vanishes
due to the finite DOS at the chemical potential, and the Mott-Hubbard phase.
Actually,
from inspection of the ground state energy we find that the Abrikosov-Suhl
resonance is always favoured when $T\to0$ and $U$ large. We
emphasize that this behaviour is strongly related to the fact that for
$d=\infty$ one always
must have $A(0)>0$, which leads to a small Kondo-effect that
eventually kills the Mott-Hubbard transition as $T\to0$. Without symmetry
breaking due to a magnetic phase transition we expect a similar situation in
finite dimensions and $T>0$ \cite{pr90}. However, a smaller value of $U_c$
as $T\to0$
must be anticipated in this case, because $A(0)$ should decrease exponentially
with temperature for sufficient large $U$.

In conclusion, we have presented a set of self-consistent equations
for the Hubbard model which allows one to calculate the properties of
strongly correlated systems in the limit of infinite dimensions.
Using a quantum Monte Carlo procedure to solve these equations we have
shown that the model
displays the expected antiferromagnetism when half-filled, and that
the single-particle density of states displays a correlation pseudogap.
The importance of this result is that it allows an essentially exact
solution of the $d=\infty$ Hubbard model in the thermodynamic limit.
Thus, solutions now exist for the model in two limits, $d=1$
and $d=\infty$. With respect to the various phenomenological models
it is surely noteworthy that the feature found at the Fermi level provides
to some extent a natural way to obtain unusual properties.
Finally, while this method is discussed in the context of
the Hubbard model, it could be applied to other strongly correlated models
(i.e. the periodic Anderson model) just by changing the form of the undressed
Green's functions in \eqs{5}{9}. Work on this problem is in progress.

We would like to acknowledge useful conversations with
D.L.\ Cox,
N.\ Grewe,
V.\ Jani\v{s},
H.\ R.\ Krishnamurthy,
M.\ Ma,
A.\ Millis,
A.\ Ol\'es,
R.\ Scalettar,
and Mathew Steiner.
One of us (TP) would
also like to thank for the hospitality and motivating atmosphere at the
physics department of the Ohio State University.
This work was supported by the National Science Foundation
grant number DMR-9107563, by the National Science Foundation grant number
DMR-88357341, the Ohio State University center of materials research
and by the Ohio Supercomputing Center.

\pagebreak


\begin{thebibliography}{99}
\bibitem{c1} The undressed Green's function occurring in the perturbation
expansion of the Anderson model is the local component of the solution of the
corresponding resonant level model, i.e.\ where $U=0$. This has the general
form $(G_l(z))^{-1}=z-\epsilon_l+\Delta(z)$.
\bibitem{comment3} Actually, from the mean-field like nature of the solution
in $d=\infty$ one would expect the $d=3$ data rather to be smaller. This should
hold, however, only if we calculated our results with a free DOS as it occurs
in a three dimensional system instead of the Gaussian (2) used here.
\end{thebibliography}
\end{document}